\documentclass{article}

\usepackage{arxiv}

\usepackage[utf8]{inputenc} 
\usepackage[T1]{fontenc}    
\usepackage{hyperref}       
\usepackage{url}            
\usepackage{booktabs}       
\usepackage{amsfonts}       
\usepackage{nicefrac}       
\usepackage{microtype}      
\usepackage{lipsum}		
\usepackage{graphicx}
\usepackage{doi}
\usepackage{placeins}

\usepackage{siunitx}
\usepackage[english]{babel}
\usepackage{amsmath,amssymb,amstext}
\usepackage{color}
\usepackage[backend=biber, style=nature, url=false,isbn=false]{biblatex}
\DeclareSourcemap{
	\maps[datatype=bibtex]{
		\map{
			\step[fieldset=abstract, null]
		}
	}
}

\let\svthefootnote\thefootnote
\newcommand\freefootnote[1]{%
	\let\thefootnote\relax%
	\footnotetext{#1}%
	\let\thefootnote\svthefootnote%
}

\title{Dynamics of wind turbine operational states}

\author{
	Henrik M. Bette$^{1,*}$, Christian Wiedemann$^2$, Matthias Wächter$^2$, Jan A. Freund$^2$, Joachim Peinke$^2$, Thomas Guhr$^1$\\
	\and
	$^1$\\
	Fakultät für Physik\\
	Universität Duisburg-Essen\\
	Duisburg, Germany \\
	\and
	$^2$ \\
	Fakultät V\\
	Carl von Ossietzky Universität\\
	Oldenburg, Germany\\
}

\renewcommand{\shorttitle}{Dynamics of wind turbine operational states}

\hypersetup{
	pdftitle={Dynamics of wind turbine operational states},
	pdfsubject={},
	pdfauthor={Henrik M.~Bette, Christian~Philipp},
	pdfkeywords={},
}

\begin{document}

\maketitle

\begin{abstract}

	Modern wind turbines gather a wealth of data with Supervisory Control And Data Acquisition (SCADA) systems. We study the short-term mutual dependencies of a variety of observables by evaluating Pearson correlation matrices on a moving time window. Using clustering on these matrices, we identify multiple stable operational states, which characterize the non-stationarity of mutual dependencies at a single turbine. They represent different turbine operational settings.
		
	Moreover, we combine the clustering analysis with a construction of a stochastic process to study the switching dynamics of those states in more detail. Calculating the distances between correlation matrices we obtain a time series that describes the behavior of the complex system in a collective way. Assuming this time series to be governed by a Langevin equation, we estimate the deterministic (drift) and stochastic (diffusion) components of the dynamics to understand the underlying non-stationarity. After adapting our method to specific features of our data, we are able to study the dynamics of operational states and their transitions as well as to resolve hysteresis effects.

\end{abstract}

\section{Introduction}

\freefootnote{*henrik.bette@uni-due.de}

The Global Wind Energy Council reports in 2023 a total installed wind capacity of 906 GW \cite{Hutchinson2023}. The main technology for harnessing this energy are wind turbines. They are complex systems that require sophisticated control strategies to regulate their operation \cite{Schutt2014, Sayed2021}. Menezes et. al. \cite{Menezes2018} provides a general review of common control strategies. Several studies have explored the analysis of such control systems in wind turbines, including both theoretical and experimental investigations \cite{Novaes2018}. One approach to analyze wind turbine control systems is by means of modeling and simulation. For example, Pustina et al. \cite{Pustina2022} presented a nonlinear predictive model approach for power maximization and tested it in simulations using the OpenFAST environment. Another approach is the usage of optimization techniques. For example, Fernandez-Gauna et al. \cite{Fernandez-Gauna2022} proposed a control strategy for wind turbines based on a predictive machine learning model. The authors managed to combine control of the pitch system and the generator torque in one optimization problem. In addition to modeling and simulation, experimental studies have also been put forward to analyze wind turbine control systems. For example, Pöschke et al. (2022) \cite{Poeschke2022} conducted a validation study of a model based control strategy in a wind tunnel. Some control strategies are also developed and analyzed for specific types of wind turbines. For example,  Lopez-Queija et. al. \cite{Lopez-Queija2022} review state of the art control systems for floating offshore turbines.

In the present contribution we pursue a new approach. We study the dynamics of operational states based on real SCADA-data (Supervisory Control And Data Acquisition). We introduce a method based on short-term Pearson correlation matrices and Langevin analysis that two of the present authors developed for financial data \cite{Rinn2015, Stepanov2015}. In a previous study we showed that clustering Pearson correlation matrices calculated over short time periods facilitates automatic distinction between operational states of the wind turbine controller \cite{BetteNonStat}. This requires the usage of high-frequency SCADA-data. Langevin analysis is a powerful approach that uses a combination of deterministic and random processes to model the dynamics of a system. It is already used in the wind energy field to study power curve behavior \cite{wachter2011}. By employing Langevin analysis on the correlation matrices, we gain a direct way to analyze the dynamics of the control system during real operation. We test the applicability of the method by analyzing data of one year from an offshore \emph{Vestas} wind turbine.

We present the methodology in Sec. \ref{sec:Dynamics_methodology}. Here, we first explain the calculation of a correlation matrix time series in \ref{sec:Dynamics_correlationTheory} and then illustrate the Langevin analysis thereof in \ref{sec:Dynamics_langevinTheory}. Afterwards, we introduce the data set used for our analysis in Sec. \ref{sec:Dynamics_data}. We present the results from applying the introduced methodology on our dataset in Sec. \ref{sec:Dynamics_results}. Here, we will also adapt the Langevin analysis to better resolve hysteresis features observed in our dataset. In Sec. \ref{sec:Dynamics_conclusion} we discuss the applicability of the presented method to wind turbines and the findings on our data. We also present some future research possibilities.

\section{Methods}
\label{sec:Dynamics_methodology}

The SCADA system of a wind turbine measures many different observables during operation. We denote the data measured for an observable $k$ at time $t$ with $X_k(t)$, $t=1,\dots,T_{\text{end}}$, $k=1,\dots,K$. We notice that the time $t$ is given in units of the uniform sampling interval $\Delta t$. For our following calculations it is helpful to arrange the data in the $K\times T_{\text{end}}$ data matrix
\begin{equation}
	\label{eq:datamatrix}
	X = \begin{bmatrix}
		X_1(1) & \dots & X_1(T_{\text{end}}) \\
		\vdots & & \vdots \\
		X_k(1) & \ddots & X_k(T_{\text{end}}) \\
		\vdots & & \vdots \\
		X_K(1) & \dots & X_K(T_{\text{end}})
	\end{bmatrix},
\end{equation}
where the $k$'th row is the time series $X_k(t)$.

\subsection{Correlation matrix clustering}
\label{sec:Dynamics_correlationTheory}

To calculate the Pearson correlation between different observables, we have to first normalize each time series $X_k(t)$ to mean value zero and standard deviation one:
\begin{equation}
	M_k(t) = \frac{X_k(t)-\mu_k}{\sigma_k} ~~, ~~ k=1,\dots,K ~~, ~~ 1 \leq t \leq T_{\text{end}} ~~,
\end{equation}
with the sample mean value
\begin{equation}
	\mu_k = \frac{1}{T_{\text{end}}}\sum_{t=1}^{T_{\text{end}}} X_k(t)
\end{equation}
and the sample standard deviation
\begin{equation}
	\sigma_k = \sqrt{\frac{1}{T_{\text{end}}} \sum_{t=1}^{T_{\text{end}}} (X_k(t)-\mu_k)^2} ~~.
\end{equation}

The normalized $K \times T$ data matrix $M$ is defined analogously to $X$ in Eq. \eqref{eq:datamatrix}. The $K \times K$ Pearson correlation matrix is then given by
\begin{equation}
	C = \frac{1}{T_{\text{end}}} M M^{\dagger} ~~,
\end{equation}
where $M^{\dagger}$ denotes the transpose of $M$. Each matrix element $C_{ij}$ is the Pearson correlation coefficient of the observables $i$ and $j$. By definition $C$ is a real-symmetric and positive semidefinite matrix with diagonal values equal to one.

The operational conditions for any wind turbine, e.g. wind and weather, are highly non-stationary. Thus, to analyze the dynamics of the correlation structure, we must calculate the correlation matrix on a moving time window. We divide the whole time series into smaller non-overlapping intervals, referred to as epochs. The length of such an epoch is chosen as $T=30$min. This value is derived from a compromise: $T$ should be as short as possible to resolve the non-stationary dynamics, but must be of a certain length so that the correlations are not dominated by sampling noise. Each epoch is labeled with the time $\tau$ and contains the measurements for $\tau \leq t < \tau + T$. The steps for the calculation of the correlation matrix are then carried out separately for each epoch data matrix $X(\tau)$ to gain a time series of correlation matrices $C(\tau)$, $1 \leq \tau \leq T_{\text{end}}-T$. We define
\begin{equation}
	\label{eq:distancemeasure}
	d(\tau, \tau') = \sqrt{\sum_{i,j}(C_{ij}(\tau)-C_{ij}(\tau'))^2} = ||C(\tau) - C(\tau')||
\end{equation}
analogous to the Euclidean distance as a distance measure between correlation matrices. On these distances a bisecting $k$-means algorithm is used to find clusters of similar correlation structures \cite{BetteNonStat}. For each cluster $s$ we then calculate the center as an element-wise mean:
\begin{equation}
	\label{eq:centermatrix}
	\langle C_{ij} \rangle_s = \frac{1}{|z_s|}\sum_{\tau \in z_s} C_{ij}
\end{equation}
with $|z_s|$ being the number of elements in cluster $s$. The center matrix is denoted $C_s$. For each cluster, we define a dynamical observable describing the total current correlation structure, i.e. operational state, of the turbine as 
\begin{equation}
	\label{eq:dynamic_distance}
	d_s(\tau) = ||C(\tau) - C_s|| ~~.
\end{equation} It is the distance between the matrix for epoch $\tau$ (current operational state) and the center for cluster $s$ (operational state expected for a given cluster $s$).

\subsection{Langevin analysis}
\label{sec:Dynamics_langevinTheory}

To analyze the time series $d_s(\tau)$ we utilize Langevin analysis. We assume that the time evolution of $d_s(\tau)$ is described by
\begin{align}
	\left. \frac{\mathrm{d}}{\mathrm{d}\tau} d_s \right\vert_{x(\tau)=x} = D^{(1)}(d_s, x) + \sqrt{D^{(2)}(d_s, x)} \cdot \Gamma(\tau),
\end{align}
with conditioning on an arbitrary additional observable $x$. The first Kramers-Moyal coefficient is the drift, denoted $D^{(1)}(d_s, x)$, and the second Kramers-Moyal coefficient the diffusion, denoted $D^{(2)}(d_s, x)$. Additionally, we introduce a delta-correlated Gaussian noise $\Gamma(\tau)$ with a variance of two. We focus on the drift estimation for $d_s$ conditioned on the placeholder $x$ (representing the time and wind observables in later analysis). If we had an infinite number of measurements at every point ($d_s, x$) and any possible time increment $\vartheta$, we would be able to compute the related increments
\begin{align}
	\Delta_{\vartheta} d_s(\tau) = d_s(\tau + \vartheta) - d_s(\tau) . \label{eq:delta_d_s_analytical}
\end{align}
and use these to calculate the $n$'th order conditional moment
\begin{align}
	\mathcal{M}^{(n)}(d_s, x, \vartheta) &= \langle \left( \Delta_{\vartheta} d_s(\tau) \right)^n\rangle |_{d_s(\tau) = d_s, x(\tau) = x}, \label{eq:conditional_moment_analytical}
\end{align}
where $\langle . \rangle$ indicates the expectation value over all measurements at this point \cite{risken1996fokker,tabar2019analysis}. Then, the $n$'th Kramers-Moyal coefficient would be given by
\begin{align}
	D^{(n)}(d_s, x) &= \lim_{\vartheta \rightarrow 0} \frac{\mathcal{M}^{(n)}(d_s, x, \vartheta)}{n! \cdot \vartheta}. \label{eq:kmc_analytical}
\end{align} 

However, the dataset retrieved from the correlation analysis described in Sec. \ref{sec:Dynamics_correlationTheory} only consists of a finite number of equidistantly sampled data points with a sample interval of $T$. We use this as the smallest possible time step $\vartheta = T$ for the stochastic analysis. Moreover, we define $\vartheta_q = q \cdot \vartheta$, where $q =1, \dots, Q$, allowing us to compute the increments of the observable $d_s$ over a time lag $\vartheta_q$ as
\begin{align}
	\Delta_{\vartheta_q} d_s(\tau) = d_s(\tau + \vartheta_q) - d_s(\tau) . \label{eq:delta_d_s_numerical}
\end{align}
Then, with our finite number of points ($d_s, x$), we employ the Nadaraya-Watson \cite{Nadaraya1964, Watson1964} estimator to approximate the $n$'th conditional moment $\hat{\mathcal{M}}^{(n)}(d_s, x, \vartheta_q)$ at the point $(d_s, x)$ over a time lag $\vartheta_q$. We use the hat to indicate quantities estimated from data.
\begin{align}
	\hat{\mathcal{M}}^{(n)}(d_s, x, \vartheta_q) &= \sum_{\tau=1}^{T_{\text{end}}-\vartheta_{q}} \left(\Delta_{\vartheta_q} d_s(\tau) \right)^n \cdot \frac{\kappa_{a, 
			b}\left(\frac{d_s(\tau)-d_s}{h_d}, \frac{x(\tau)-x}{h_x} \right)}{\sum\limits_{\tau=1}^{T_{\text{end}}-\vartheta_{q}}\kappa_{a, 
			b}\left(\frac{d_s(\tau)-d_s}{h_d}, \frac{x(\tau)-x}{h_x} \right)} \label{eq:conditional_moment_numerical_1}
\end{align}
Here, $\kappa_{a, b}(y_1, y_2)$ represents a two-dimensional kernel, and $h_d$ and $h_x$ correspond to the bandwidths utilized for the estimation. We calculate the two-dimensional kernel as the product of two one-dimensional kernels, i.e. a D-kernel,
\begin{align}
	\kappa_{a, b}(y_1, y_2) = \kappa_a(y_1) \cdot \kappa_b(y_2) .
\end{align}
For the one-dimensional kernels we employ a Gaussian form \cite{silverman1986, Wied2012}
\begin{align}
	\kappa_{\text{G}}(y) = e^{-\frac{1}{2} y^2} .
\end{align}
The bandwidth associated with the kernel function is just as crucial as the kernel function itself. When examining large-scale structures, it is useful to employ wider bandwidths. However, larger bandwidths may hamper resolution of small-scale structures. For the analysis of our data we found that the following values give good results and used them unless stated otherwise:
\begin{itemize}
	\item for wind speed $h_{\text{WindSpeed}} = 0.5$m/s
	\item for time $h_t = 1.6$h
	\item for the inter matrix distance $\displaystyle h_{d_s} = \frac{\max\limits_{\tau}(d_s(\tau))}{30}$.
\end{itemize}

We make the assumption that for small time increments $\vartheta_q$, the conditional moments $\hat{\mathcal{M}}^{(n)}(d_s, x, \vartheta_q)$ exhibit linearity, and there is no additional measurement noise, implying $\hat{\mathcal{M}}^{(n)}(d_s, x, 0) = 0$ \cite{Siefert2003}. By averaging the conditional moments divided by the employed time increment $\vartheta_q$, as depicted in Eq. \eqref{eq:kmc_numerical}, we estimate the $n$-th Kramers-Moyal coefficients based on the provided estimations of the $n$-th conditional moments: \cite{risken1996fokker,tabar2019analysis}
\begin{align}
	\hat{D}^{(n)}(d_s, x) &= \frac{1}{Q} \sum_{q}^{Q} \frac{\hat{\mathcal{M}}^{(n)}(d_s, x, \vartheta_q)}{n! \cdot \vartheta_q} \label{eq:kmc_numerical}
\end{align}
As the minimal time step $\vartheta$ already is $30$ minutes, we only consider $Q=1$, so that $\vartheta_q$ will not be too large. Here, again a shorter epoch length $T=\vartheta$ would be desirable, but is not feasible due to the correlation matrix calculation.

Sometimes a more intuitive description is the potential
\begin{align}
	\hat{\Phi}(d_s, x) = - \int \hat{D}^{(1)}(d_s, x) \text{d} d_s
\end{align}
calculated from the first Kramers-Moyal coefficient. Local minima of the potential correspond to stable fixed points in the system. When looking at the drift, which is the derivative of the potential, these points are indicated by values of zero and a negative slope at the zero crossing. Throughout our analysis, we will use whichever is best suited for understanding a particular issue.

\section{Data Set}
\label{sec:Dynamics_data}

The data we use stems from the Supervisory Control and Data Acquisition (SCADA) system of a \textsc{Vestas} turbine. It is situated in an offshore wind farm off the coast of Great Britain. The data were measured approximately every 5s for the year 2017. To obtain consistent time stamps and a stable frequency, i.e. a consistent sampling interval $\Delta t$, the data were aggregated on 10s time intervals by averaging. If no data were measured in the original 5s set during a 10s time interval, then there will also be missing data in our aggregated set.

The analyzed data contains measurements of five observables:
\begin{itemize}
	\item generated active power (ActivePower)
	\item generated current (CurrentL1)
	\item rotation per minute of the rotor (RotorRPM)
	\item rotation per minute of the high speed shaft at the generator (GeneratorRPM)
	\item wind speed (WindSpeed)
\end{itemize}
As there are no deviations between the three phases of the generated current, we simply choose one of them. We expect from the \textsc{Vestas} turbine a control shift from a low wind speed regime with variable rotation speed to an intermediate regime with constant rotation to a rated region with constant rotation and produced power. The studied observables are suitable to analyze these changes. The two rotational speed observables as well as current and active power are usually strongly coupled observables. We include all of them in our study as group correlations are important for the characterization of correlation structures. When trying to detect anomalies, for example, such structures might break up. It is also consistent with the first study identifying operational states based on the correlation matrix \cite{BetteNonStat}.

\section{Results}
\label{sec:Dynamics_results}

\subsection{Correlation matrix states}
\label{sec:Dynamics_results_states}

\begin{figure}
	\centering
	\includegraphics[width=0.95\textwidth]{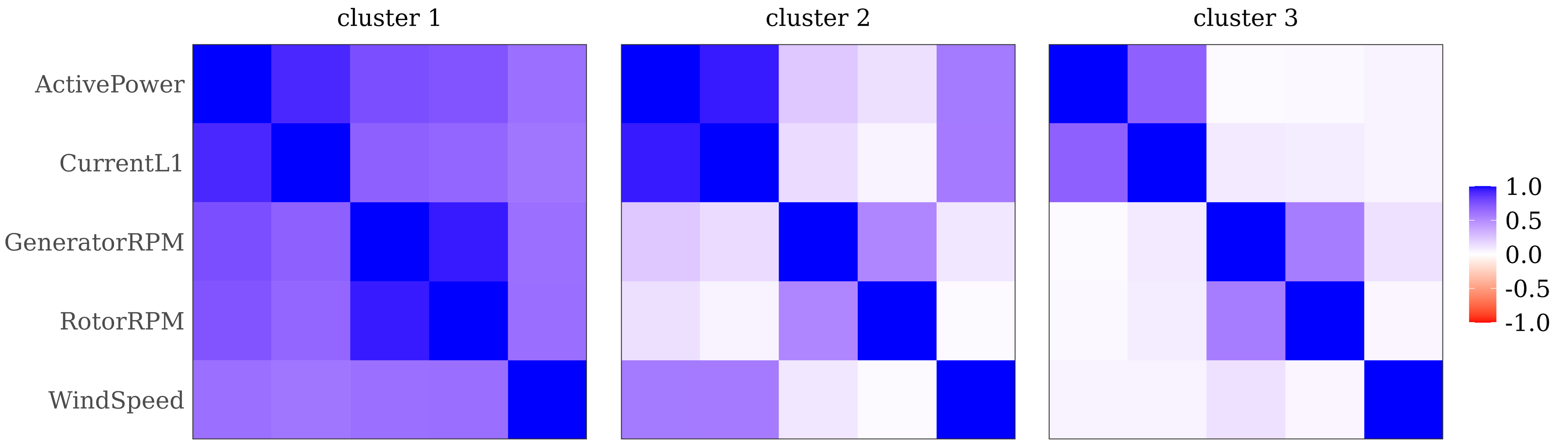}
	\caption{Cluster centers of Pearson correlation matrices calculated according to Eq. \eqref{eq:centermatrix}. The x- and y-axes are identical and display the different observables, for presentation we only labeled the y-axis. The matrix entries are represented by color.}
	\label{fig:clusterCenters}
\end{figure}

As a first step to analyze the operational dynamics of wind turbines, we must automatically distinguish the different operational states of the wind turbine. We apply a method developed by two of the present authors for this in \cite{BetteNonStat}. For each epoch, we calculate the Pearson correlation matrix. The resulting set may be viewed as a time series of matrices. After applying clustering to this set, we obtain the three cluster centers shown in Fig. \ref{fig:clusterCenters} according to Eq. \eqref{eq:centermatrix}. Their number is decided based on visual inspection as well as the silhouette coefficient \cite{BetteNonStat}. They represent different operational states of the turbine: At low wind speeds exists a regime where stronger winds lead to faster rotation, which in turn leads to more generated power. This is represented by Cluster 1. Intermediate wind speeds are best used by keeping the rotation at a constant, optimal value. This is indicated by the vanishing correlations of the rotational observables in Cluster 2. Here, more power is generated by increasing the torque. For high wind speeds, the turbine operates at rated power output, i.e. it has reached its upper power production limit. This is seen in Cluster 3. Rotation and produced power are both decoupled from the wind speed. The clustering procedure and wind speed distributions for each cluster are presented in detail in \cite{BetteNonStat}. When choosing different observables, the clusters and possibly their number will change.

\begin{figure}
	\centering
	\includegraphics[width=0.8\textwidth]{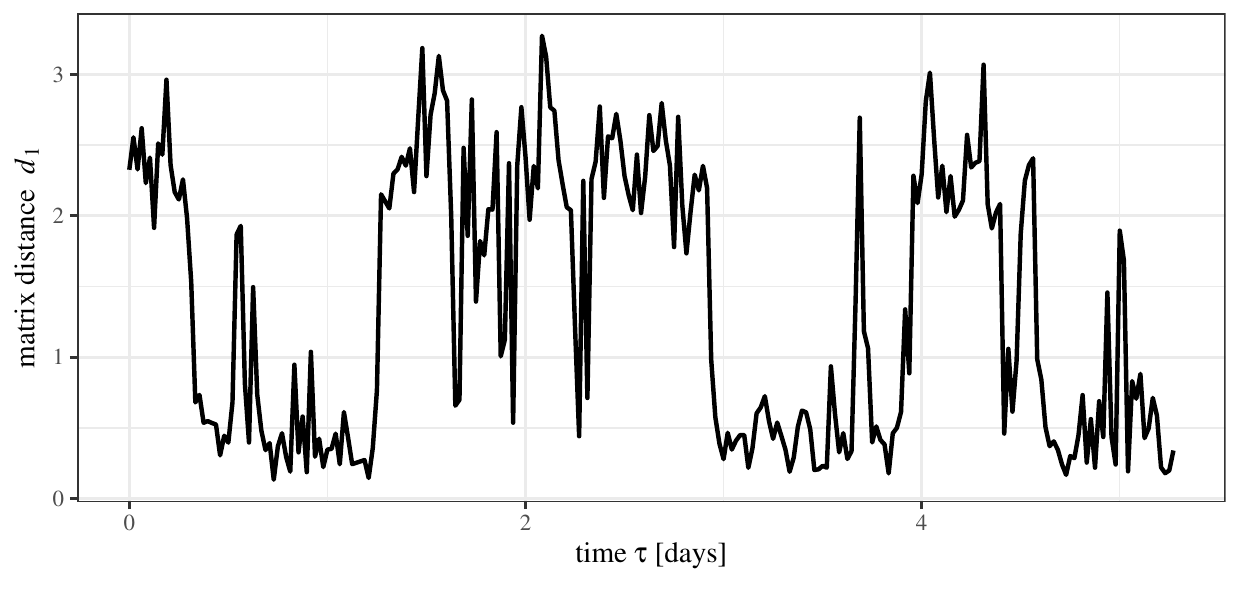}
	\caption{Matrix distance $d_1$ versus time $\tau$ for an arbitrary time span.}
	\label{fig:dynamic_distance}
\end{figure}

The cluster centers represent typical operational states. However, with switching between different states and constantly changing external conditions, the correlation matrix for any epoch will usually fluctuate around the identified states. The current correlation structure is described in an aggregated way by the distance between the current correlation matrix and a chosen cluster center as calculated in Eq. \eqref{eq:dynamic_distance}. Thereby, we effectively look at the system from the viewpoint of one cluster center. An example time span for this matrix distance measure is shown in Fig. \ref{fig:dynamic_distance}. We only plot $d_1$ as an example. This time series appears to be a stochastic process with noise fluctuating around fixed points (the cluster centers). Therefore, we now try to extract the deterministic components of its behavior to study the switching dynamics of the operational states.

\subsection{Dynamics versus Time}
\label{sec:Dynamics_results_time}

\begin{figure}
	\centering
	\includegraphics[width=0.95\textwidth]{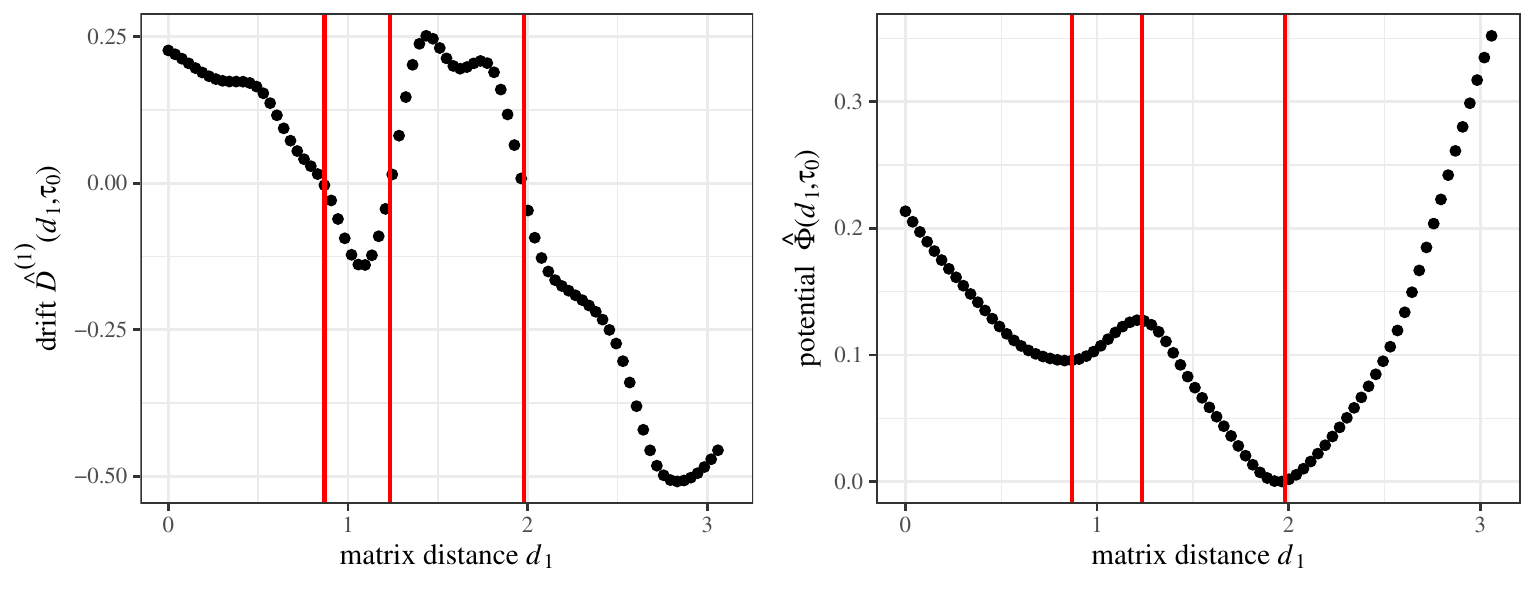}
	\caption{Drift $\hat{D}^{(1)}(d_s, t_0)$ and corresponding potential $\hat{\Phi}(d_1, t_0)$ versus matrix distance $d_1$ at time $\tau=\tau_0$. Here, the bandwidth for the distance is chosen as $\displaystyle h_{d_1} = \max\limits_{\tau}(d_1(\tau))/10$. Red lines approximately indicate the values of $d_1$ where the drift crosses zero and thus stable or unstable fixed points, also seen in the potential.}
	\label{fig:pot2D}
\end{figure}

Combining the identified clusters with the Langevin analysis as described in Sec. \ref{sec:Dynamics_methodology}, we are now able to analyze the dynamics of operational states and their transitions. First, we look at the dynamics versus time, i.e. $x\widehat{=}\tau$. At a fixed point in time $\tau=\tau_0$, we obtain a drift $\hat{D}^{(1)}(d_s, \tau_0)$ and corresponding potential $\hat{\Phi}(d_s, \tau_0)$ versus matrix distance $d_s$. As an example, we show this for the matrix distance $d_1$ to Cluster 1 in Fig. \ref{fig:pot2D}. Any positive value in the drift means that the system tends to move to larger $d_1$, a negative value means the opposite. Therefore, a zero crossing in the drift with negative slope indicates a stable fixed point. In the potential this is represented as a minimum. If the drift crosses zero, but with a positive slope, the potential has a local maximum indicating, in general, an unstable fixed point. Here, one deep minimum exists representing the dominating operational state at $\tau=\tau_0$. The second, less deep, minimum indicates the presence of a second operational state in the vicinity (cf. the bandwidth in Eq. \eqref{eq:conditional_moment_numerical_1}) of $t=t_0$.

\begin{figure}
	\centering
	\includegraphics[width=0.80\textwidth]{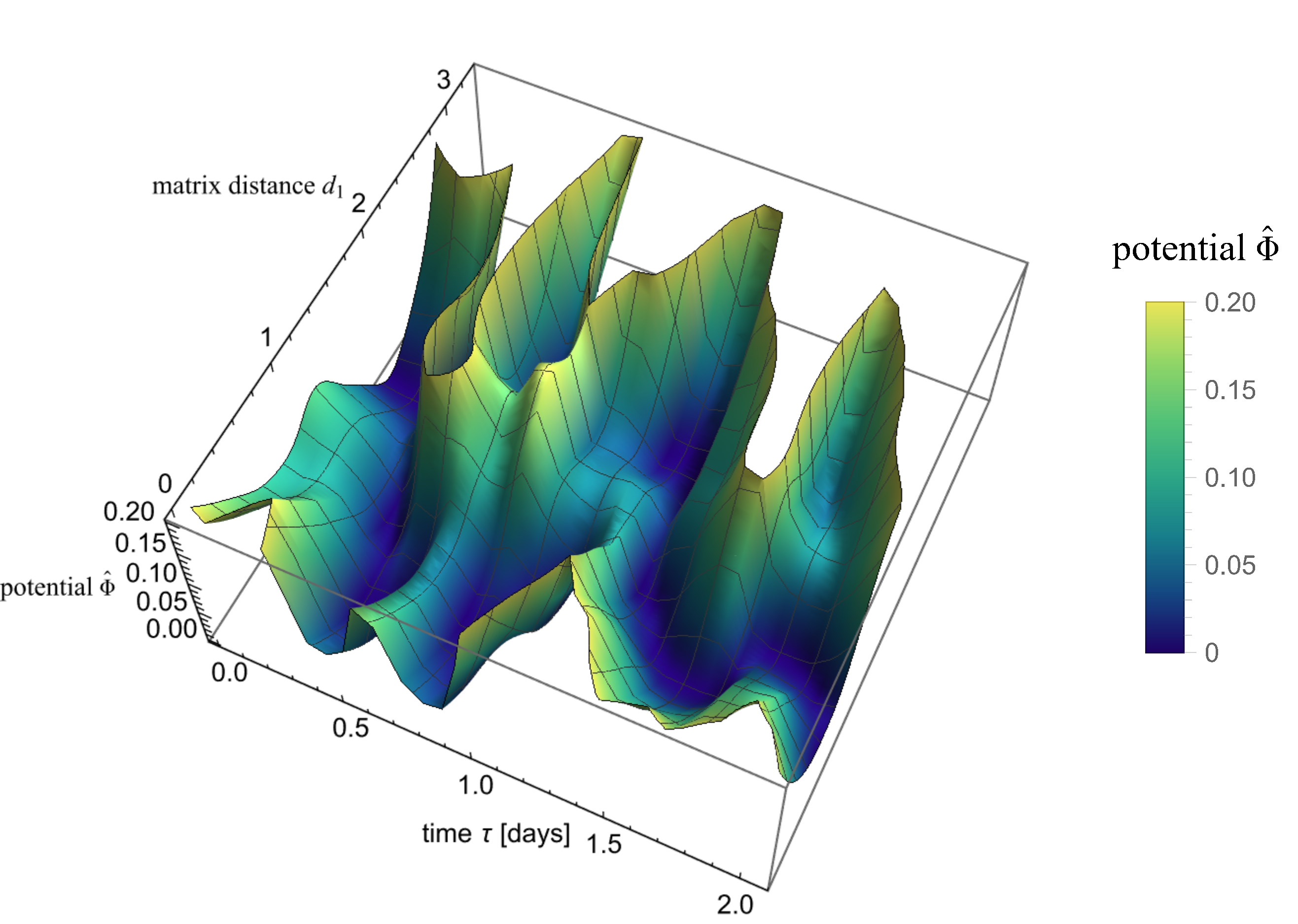}
	\caption{Potential $\hat{\Phi}(d_1, \tau)$ versus matrix distance $d_1$ and time $\tau$. The shown total time span is approximately two days. The range of shown values is restricted for increased readability. Here, the bandwidth for the distance is chosen as $\displaystyle h_{d_1} = \max\limits_{\tau}(d_1(\tau))/10$.}
	\label{fig:potentialOverTime}
\end{figure}

Now, we look at the non-stationarity over time. In Fig. \ref{fig:potentialOverTime} the potential is shown for a time span of two days. We see that it changes quite quickly over time. As expected the correlation matrix, i.e. the operating state, is not stable over time and therefore the potential changes. It is clear that the transitions between the states happen quite often (here, about 5 times in two days). However, we see that for the short periods where no changes happen, a clear minimum in the potential exists, indicating stability during these periods.

\subsection{Dynamcis versus Wind Speed with Standard Drift Estimation}
\label{sec:Dynamics_results_windspeed}

\begin{figure}
	\centering
	\includegraphics[width=0.99\textwidth]{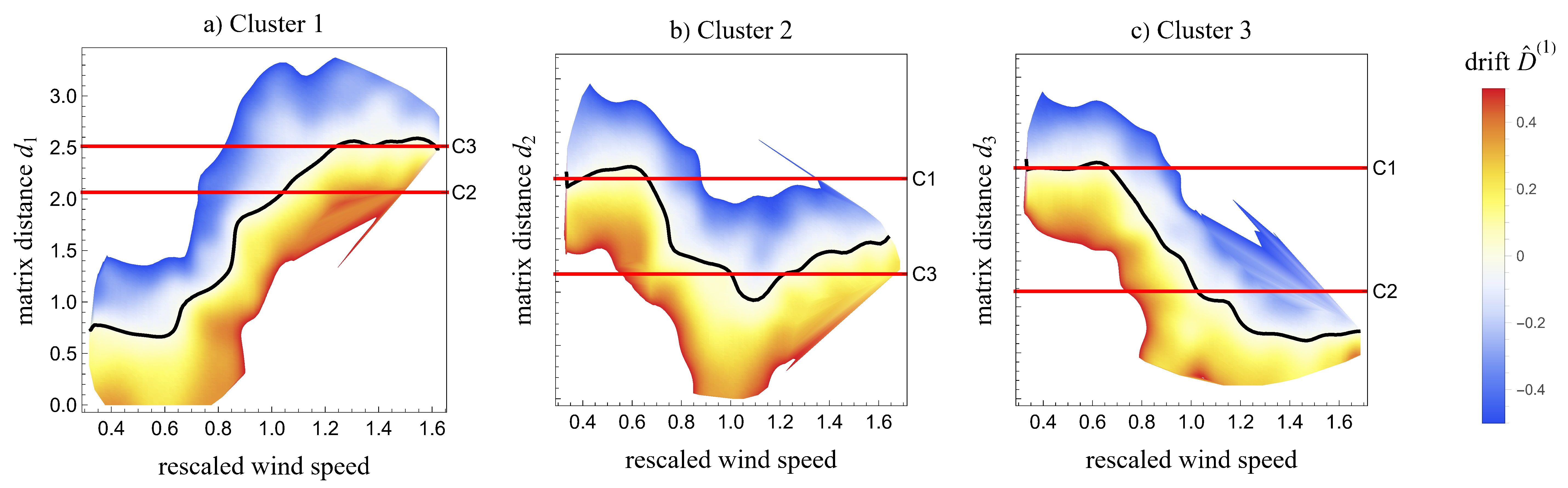}
	\caption{Color profile of drift $\hat{D}^{(1)}(d_s, u)$ versus matrix distance $d_s$ and rescaled wind speed. For panels a), b) and c) the cluster $s$ is chosen to be 1, 2, and 3, respectively. The conditional moments $\hat{\mathcal{M}}^{(1)}(d_s, u, \vartheta_q)$ for the drift estimation were calculated according to Eq. \eqref{eq:conditional_moment_numerical_1}. The range of shown values is restricted for increased readability. The black line indicates where the drift is zero. Red lines indicate the distance to the other two cluster centers.}
	\label{fig:driftAll}
\end{figure}

\begin{figure}
	\centering
	\includegraphics[width=0.99\textwidth]{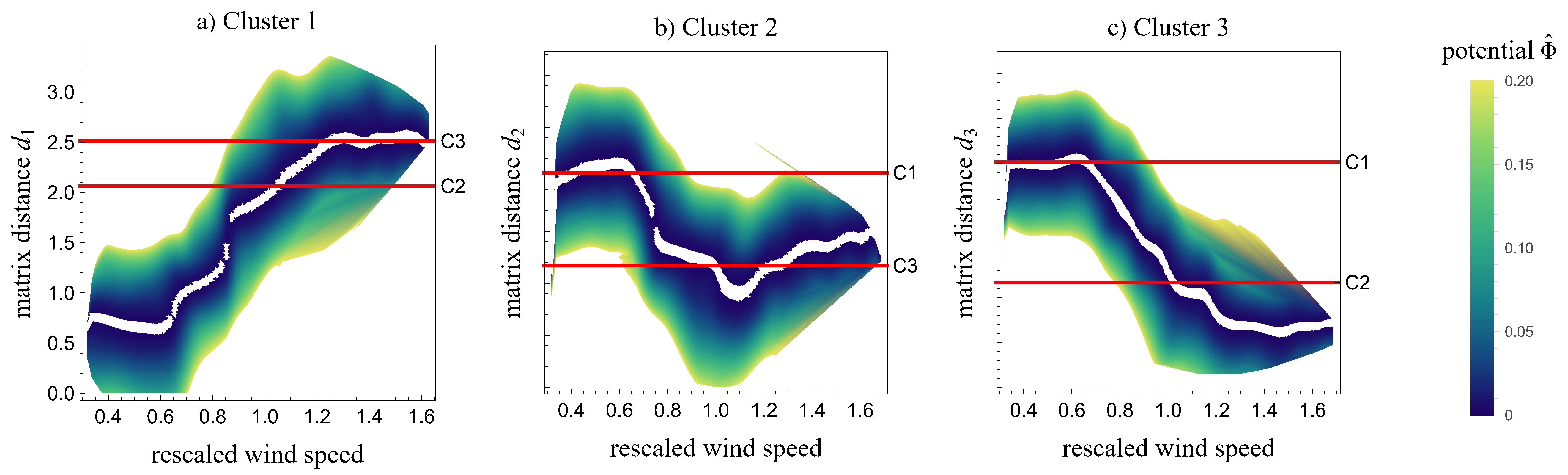}
	\caption{Color profile of potential $\hat{\Phi}(d_s, u)$ versus matrix distance $d_s$ and rescaled wind speed. For panels a), b) and c) the cluster $s$ is chosen to be 1, 2, and 3, respectively. The conditional moments $\hat{\mathcal{M}}^{(1)}(d_s, u, \vartheta_q)$ for the drift and potential estimation were calculated according to Eq. \eqref{eq:conditional_moment_numerical_1}. The range of shown values is restricted for increased readability. The white area indicates the minimum of the potential, i.e. $\hat{\Phi}(d_s, u)<0.0005$. Red lines indicate the distance to the other two cluster centers.}
	\label{fig:potAll}
\end{figure}

In \cite{BetteNonStat} we found that the operational states primarily depend on wind speed. Hence, another way to look at our system is by studying the non-stationarity not versus time, but versus wind speed, i.e. $x\widehat{=}u$. Thereby, we do not find the quick changes due to environmental conditions any longer. We rather see the transitions between operating states as they change with wind speed. Figure \ref{fig:driftAll} shows the drift as viewed from cluster centers 1, 2 and 3. In Fig. \ref{fig:potAll} we also show the corresponding potentials for comparison. Essentially, at all times the drift is zero and the crossing occurs with negative slope, we find a minimum in the potential. These are the points where the system is stable. For a given wind speed, the system usually drifts to the matrix distance where the drift is zero. As the matrices in each cluster still fluctuate around their center, the stable fixed point for a cluster $s$ when analyzing the matrix distance $d_s$ in relation to that same cluster does not lie at zero, but at small positive values.


While the zero values of the drift do not exactly coincide with the distances between clusters at all times, the qualitative behavior is generally as expected. Clusters 1, 2 and 3 develop from low to high wind speeds. Cluster 1 dominates for rescaled wind speeds (RWS) between 0.3 and 0.7. Cluster 2 is the system's desired state between 0.85 and 1.15 on the RWS axis as is Cluster 3 between RWS=1.05 and RWS=1.6. We see, that Cluster 2 and 3 share the interval from RWS=1.05 to RWS=1.15. Here, the transition between these two clusters is visible. Clusters 1 and 2 do not seem to overlap. Their transition appears more complicated. When viewed from Cluster 1 in Fig. \ref{fig:driftAll} a) it seems, there might be another, not yet defined state in the region from 0.6 to 0.9 on the RWS axis. It appears close to Cluster 1 in the matrix distance values, but shows a rather sharp transition into Cluster 2 at around RWS=0.9. However, viewed from Cluster 2 in Fig. \ref{fig:driftAll} b) it appears to be the other way around. The intermediate state is close to Cluster 2 and shows a sharp transition to Cluster 1 at around RWS=0.7. Here, the state seems to also exist for higher RWS up to 1. Viewed from Cluster 3 in Fig. \ref{fig:driftAll} c) one is inclined to infer two intermediate states between Cluster 1 and 2. Using the clustering algorithm to distinguish more than three states does not reveal such intermediate states, rather, smaller clusters of outliers with very few matrix elements split off Cluster 3.

\begin{figure}
	\centering
	\includegraphics[width=0.7\textwidth]{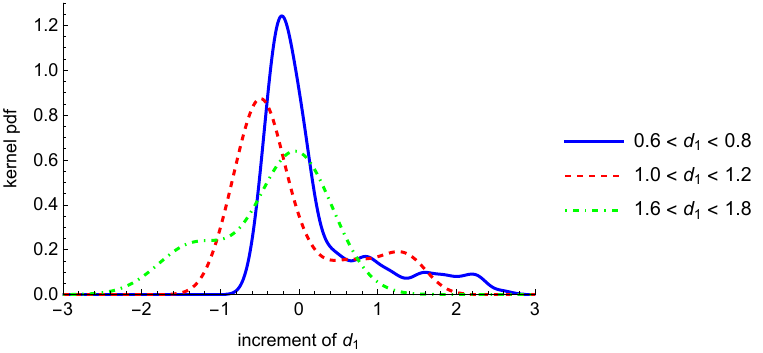}
	\caption{Kernel probability density functions (pdf) of increments in matrix distance $d_1$ measured in the data. The distributions are calculated for the RWS area: $0.65<\text{RWS}<0.75$. The covered area of matrix distance $d_1$ is indicated by color.}
	\label{fig:pdfs}
	\end{figure}

One possible explanation is the occurrence of changing environmental conditions during the 30 min intervals used for the calculation of the correlation matrices. This might then lead to correlation structures, which represent an average between different clusters. It is unclear if they are prominent enough to cause the appearance of a new intermediate stable point in the Langevin analysis. Another possibility is hysteresis in the control behavior, i.e. depending on previous conditions the controller does not always choose for the same operational conditions the same system behavior. Coming from low wind speeds, the turbine does not switch from State 1 to State 2 quickly, rather only if higher wind speeds persist for a certain amount of time. It is the other way for a transition from higher to lower wind speeds. This might lead to two fixed points for the same wind regime. Usually, this is resolved in Langevin analysis. However, in our case these two fixed points do not coexist at the same time. At any given time, depending on the operational state and its hysteresis, only one fixed point is present. If so, in the averaging process over the single increments in the data as calculated in Eq. \eqref{eq:conditional_moment_numerical_1} these two different behaviors are mixed. That this is true - at least to a certain extent - is seen in Fig. \ref{fig:pdfs}. We show three smooth kernel probability density functions for the increments in different regions of the RWS and matrix distance space. The overlap of different drift behaviors is obvious. In Fig. \ref{fig:pdfs} we see for small matrix distances $d_1$ (blue curve) a large peak at approximately zero, which stems from the fixed point for Cluster 1. However, large values for the increments exist also, which result from times when the controller tries to realize operational state 2. The mean of the distribution therefore lies at small positive values as seen in Figs. \ref{fig:driftAll} a) and \ref{fig:potAll} a). The distribution in Fig. \ref{fig:pdfs} for large $d_1$ (green curve) shows the same effect, but the other way around. For intermediate $d_1$ (red curve) we see positive and negative values of the increments in Fig. \ref{fig:pdfs}, which lead to an average close to zero. Effectively, each of these probability densities is an overlap of two different densities stemming from the two different clusters.

\subsection{Dynamcis versus Wind Speed with Modified Drift Estimation}
\label{sec:Dynamics_results_windspeed2}

\begin{figure}
	\centering
	\includegraphics[width=0.99\textwidth]{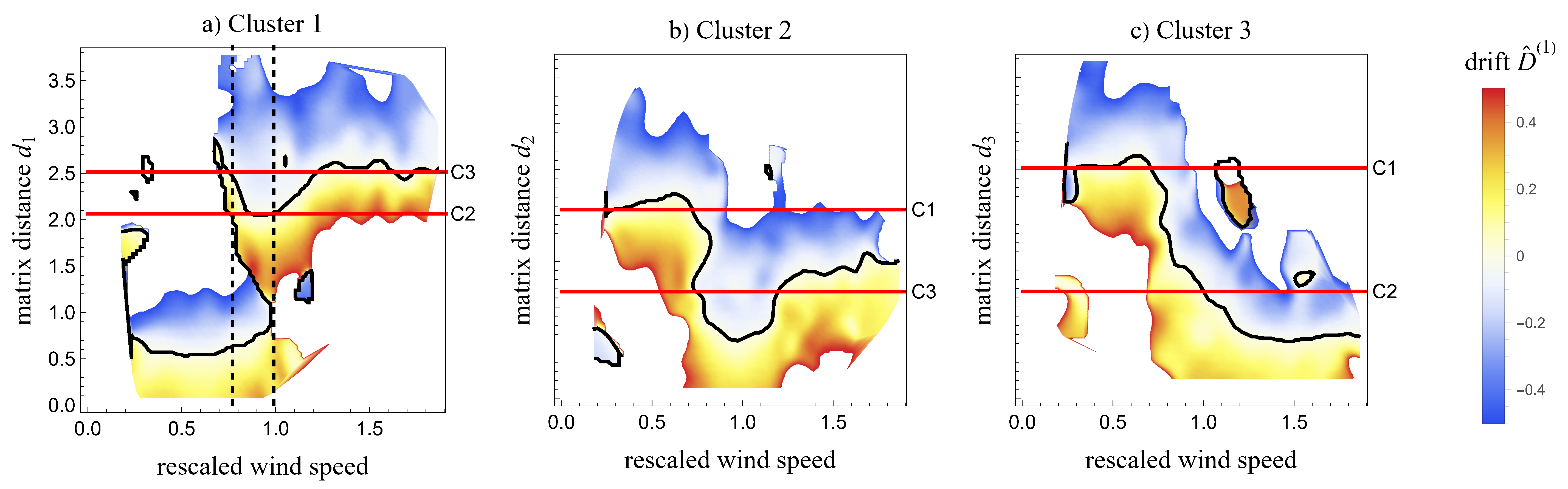}
	\caption{Color profile of drift $\hat{D}^{(1)}(d_s, u)$ versus matrix distance $d_s$ and rescaled wind speed. For panels a), b) and c) the cluster $s$ is chosen to be 1, 2, and 3, respectively. The conditional moments $\hat{\mathcal{M}}^{(1)}(d_s, u, \vartheta_q)$ for the drift estimation were calculated according to our newly proposed peak determination. The range of shown values is restricted for increased readability. The black line indicates where the drift is zero. Red lines indicate the distance to the other two cluster centers. Black, dashed lines indicate the bistable region in panel a) for $s=1$ between RWS=0.77 and RWS=0.99 according to Fig. \ref{fig:newPot_evolution}.}
	\label{fig:newDriftAll}
\end{figure}

To better resolve the issue of overlapping increment densities stemming from different clusters, we introduce a new estimation method for the drift. We attempt to determine the drift associated only with the density of the cluster, whose fixed point is closest in the space of matrix distance and wind speed. This cluster should be more likely to appear at this point as, in general, a system might be more frequently around its fixed point than very far away from it. Therefore, this cluster should be the origin of the highest peak in the pdf as seen in Fig. \ref{fig:pdfs}. In fact, Fig. \ref{fig:pdfs} indicates that the hysteresis might be better visible if one estimated the drift as the maximum of the shown distributions instead of the average. This means we effectively trim the densities by excluding outliers to a symmetric distribution around the highest peak. Then, mean value and highest peak occur at the same value, which should be close to the drift value associated with only one cluster. Thus, we proceed from the estimation of conditional moments in Eq. \eqref{eq:conditional_moment_numerical_1} to a new method. For the estimation at point $(d_s, x)$, we consider all increments of $d_s$ to the power $n$ over a time lag $\vartheta_q$ such that
\begin{align}
	\left(\Delta_{\vartheta_q} d_s(\tau) \right)^n, ~ \tau=1,\dots, T_{\text{end}}-\vartheta_q .
\end{align}
For these quantities, we calculate a smooth kernel probability density function considering the same Gaussian weights as before in Eq. \eqref{eq:conditional_moment_numerical_1}. We choose its maximum as the new $\hat{\mathcal{M}}^{(n)}(d_s, x, \vartheta_q)$. As we are interested in the deterministic drift, we set $n=1$. All remaining calculations to derive the drift and potential remain as before. We display results for this new estimation of the drift in Fig. \ref{fig:newDriftAll}. We see that the general behavior is similar to the standard drift estimation shown in Fig. \ref{fig:driftAll}, but we resolve more details.

At very low wind speeds, we identify a regime, where Cluster 2 appears to be dominant. This was not seen before at all in Fig. \ref{fig:driftAll}. However, for the analyzed turbine this behavior is correct. A non-vanishing probability for Cluster 2 to occur in this regime was also seen in the extended cluster analysis in \cite{BetteNonStat}. The system tries to keep the rotation constant at a minimal viable rotation frequency for very low wind speeds. In the correlation structure this looks the same as the later on appearances of Cluster 2 even if the fixed value of the rotation is different here.

\begin{figure}
	\centering
	\includegraphics[width=0.95\textwidth]{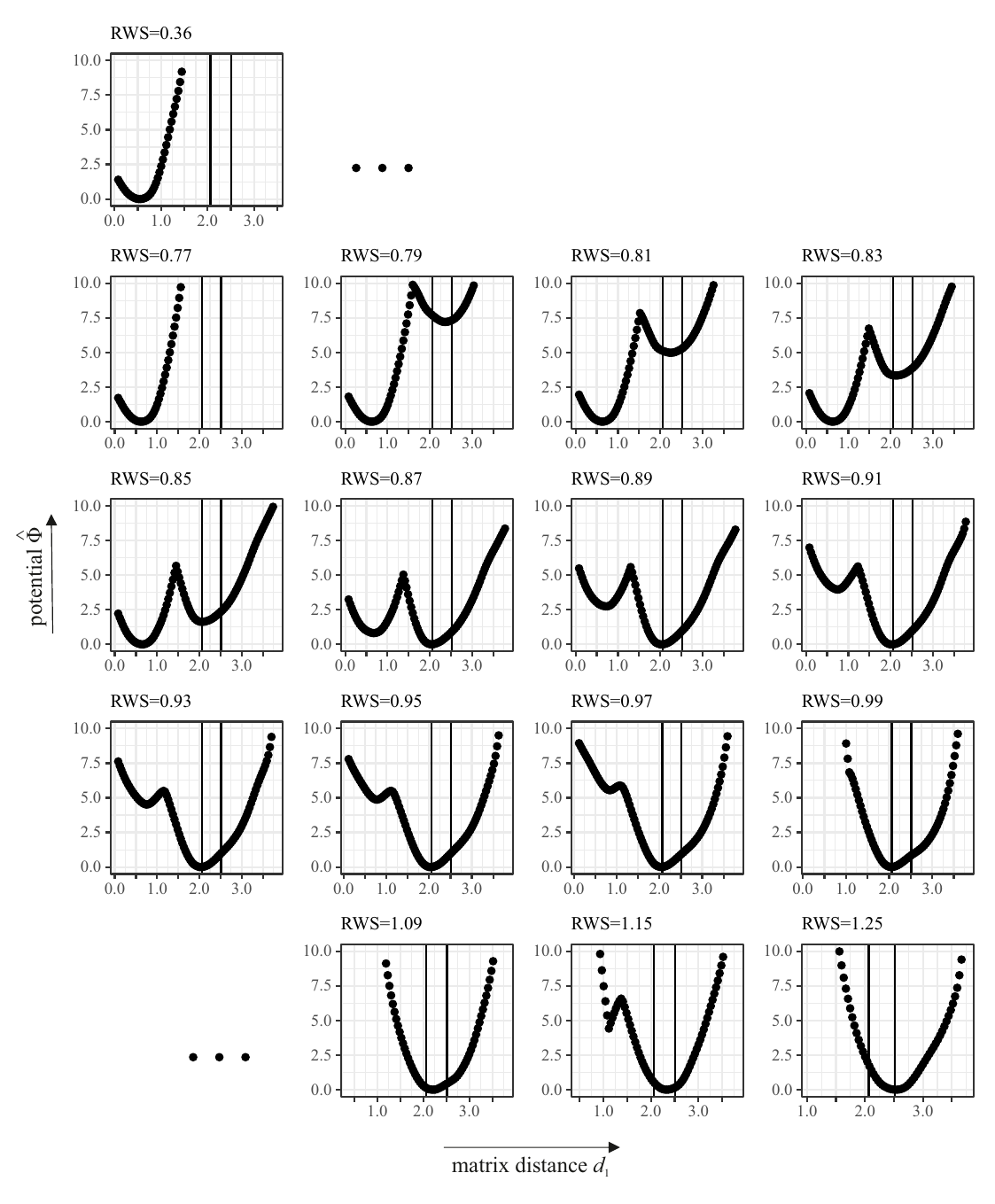}
	\caption{Potential $\hat{\Phi}(d_1, \tau)$ versus matrix distance $d_1$. Each subplot represents the potential at the indicated RWS. From upper left to lower right the RWS increases. The vertical line represents the distance to the center of Cluster 2. The individual plots are vertical slices of the potential corresponding to Fig. \ref{fig:newDriftAll} a).}
	\label{fig:newPot_evolution}
\end{figure}

For higher wind speeds, between RWS$\approx 0.3$ and RWS$\approx 0.8$ Cluster 1 is dominant. We see good alignment of the distance to the cluster center (red lines) and the zero values of the drift when viewed from Clusters 2 or 3 in Figs. \ref{fig:newDriftAll} b) and c). Viewed from Cluster 1 the drift is zero at small positive values in the matrix distance $d_1$. This is because the distance can only take positive values and the system still fluctuates around the center of the cluster. Starting from RWS$\approx 0.77$ to RWS$\approx 0.99$ we now resolve the complicated transition between Clusters 1 and 2. This is best seen when the system is viewed from Cluster 1 in Fig. \ref{fig:newDriftAll} a). Here, a bistable region exists where both clusters exist for the same wind speeds due to hysteresis. We show this transition in detail in Fig. \ref{fig:newPot_evolution}. We see that for RWS=0.77 one clear minimum exists representing Cluster 1. With increasing wind speed a second local minimum appears, which represents Cluster 2. At the beginning of transition the global minimum lies clearly at Cluster 1. For these low wind speeds State 2 is possible, but most often the turbine will be in State 1. Then, at RWS$=0.87$ both minima have approximately the same depth, before with even higher wind speeds the first minimum representing Cluster 1 starts to become less deep than the one for Cluster 2. Here, at higher wind speeds, Cluster 1 might appear when the turbine did not yet switch to Cluster 2, but it becomes ever less likely. The potential minimum associated with Cluster 1 finally vanishes at RWS$=0.99$.

Next, Cluster 2 is dominant at around RWS$=1.0$. This is best seen when viewed from Cluster 2 in Fig. \ref{fig:newDriftAll} b), but there is also a clear overlap between the zero values of the drift with the distance to Cluster 2 in Fig. \ref{fig:newDriftAll} a). It does not take much higher wind speeds for the system to start shifting to Cluster 3. Interestingly, this transition is apparently not affected by hysteresis. We see a steady shift in the drift, also found in the potentials shown in Fig. \ref{fig:newPot_evolution}. In a physics interpretation, the transitions from Cluster 1 to 2 and from 2 to 3 resemble phase transitions of first and second order, respectively. From RWS$\approx1.25$ onward, the system operates in State 3, i.e. with nominal power output. The drift values of zero and the distance to Cluster 3 coincide well, especially viewed from Cluster 1 in Fig. \ref{fig:newDriftAll} a).

Overall, the regimes seem to be best resolved when viewed from Cluster 1, where the distances to Clusters 2 and 3 match the drift values of zero. When looking at matrix distances $d_2$ and $d_3$, Cluster 1 is resolved well whereas it appears hard to resolve the other cluster. This might be due to cluster centers 2 and 3 not being very far apart and the necessity  for a non-vanishing bandwidth due to limited amounts of data. The qualitative behavior is visualized and estimated well in all cases. In the hyperspace of the correlation matrices, the different cluster centers apparently do not lie on a straight line. This means that the matrix distances $d_s$ contain different information for each $s$. For example, the hysteresis between States 1 and 2 is best seen from Cluster 1. 

\FloatBarrier

\section{Conclusions}
\label{sec:Dynamics_conclusion}

We studied the dynamics of operational states in wind turbines on the basis of real data. First we calculated Pearson correlation matrices on a moving time window, which facilitates studying the non-stationarity of wind turbine systems. Clustering the correlation matrices for one year of an offshore wind turbine, we identified three main operational states. Then, we combined our cluster analysis with Langevin analysis. The distance between a correlation matrix at any time $t$ and one of the cluster centers proved to be a good indicator of the current operational status. Hereby, we effectively reduced the multidimensional system, in our case correlations of five different observables, to a one-dimensional time series. Langevin analysis of this new time series provided information on the operational dynamics. Hence, it was possible to describe the complex dynamics of the multidimensional system in a simplified way. This is, in general, also possible for more observables and larger systems. We visualized the drift and corresponding potential for an intuitive interpretation of the system dynamics.

For our data, a \textsc{Vestas} turbine and five observables, we found three correlation matrix clusters, i.e. operational states of the wind turbine. The Langevin analysis allowed us to identify the wind regimes, where these states are stable. Furthermore, the regimes where transitions happen are also identified. While the presented method was previously applied in Econophysics \cite{Rinn2015, Stepanov2015}, we developed some adaptations of the method to the peculiarities of wind turbine data. Thereby, we succeeded in studying the nature of the state transitions. Between the states with constant rotation and operation at rated power we saw a smooth transition. The transition at lower wind speeds from a variable rotation state to constant rotation on the other hand, shows hysteresis behavior. In a physics interpretation they resemble second and first order phase transitions, respectively. We identified the existence of multiple fixed points in the low wind speed transition regime due to control hysteresis. Here, two fixed points exist in the system, but they never occur at the same time. This effectively leads to an overlap of different drift fields. With our new estimation method we were able to resolve these two fields. Thereby, we resolved the hysteresis effect as a bistable wind speed regime. Our new method is suited for analyzing transitions with hysteresis, where the potential minima never coexist at the same time.

Overall, while the control systems of a specific turbine are known at least to the manufacturer, our approach allows the analysis and visualization of their dynamics during real operation. The method is transferable to other wind turbine models and different choices of observables. Two caveats are in order: First, the observed operational states can also depend on more external factors than the wind speed. It might be necessary to study the drift and potential conditioned on multiple observables, but this does not constitute a problem in principle. Second, one has to carefully consider the number of states \cite{BetteNonStat}. Here, the Langevin analysis is helpful. We saw in the present study that it potentially reveals states that were not identified in the clustering. In Sec. \ref{sec:Dynamics_results} we saw the possibility for another state between Clusters 1 and 2, which we identified as a transition regime with hysteresis.

Apart from a direct study of non-stationary dynamics in turbine operation, our results are interesting for normal behavior modeling. This is part of another prominent research topic in wind energy: the early prediction of failures \cite{Tautz-Weinert2017, Maldonado-Correa2020}. Here, if one wants to detect anomalies, one must first define what is normal. This, of course, includes different behaviors introduced by the controller. Our analysis helps with the identification of wind speed regions where one can be sure of what is normal and transition periods where this is potentially unclear.

We showed applicability of the presented method to study the operational dynamics. Interesting extensions for future work remain. First, shorter epochs for the calculation of the correlation matrix are desirable requiring data of higher resolution. This might help to resolve changing conditions during an epoch. Thereby, the resolution of the Langevin estimation could be increased. Given the required data, transfer to different wind turbines models is possible and might facilitate comparison of systems in real operation. If data is available, it might also be interesting to study the dynamics under different environmental conditions such as onshore and offshore. Lastly, our new peak-based estimation of the drift is a stronger deviation from the theoretical description in Eq. \eqref{eq:conditional_moment_analytical} than the standard averaging. Further research on its general applicability, maybe of a more theoretical nature, would be interesting.

\section*{Acknowledgments}
We are grateful to \emph{Vattenfall AB} for providing the data. We acknowledge fruitful conversations with David Bastine, Timo Lichtenstein and Anton Heckens. This study was carried out in the project \emph{Wind farm virtual Site Assistant for O\&M decision support – advanced methods for big data analysis} (WiSAbigdata) funded by the Federal Ministry of Economics Affairs and Energy, Germany (BMWi), three of us (HMB, CW and MW) thank for financial support.


\printbibliography

\end{document}